\documentclass[12pt,preprint]{aastex}
\usepackage[usenames]{color}

\newcommand{\feka}{\hbox{Fe\,K$\alpha$}}

\newcommand{\simgt}{\lower 2pt \hbox{$\, \buildrel {\scriptstyle >}\over {\scriptstyle\sim}\,$}}
\newcommand{\simlt}{\lower 2pt \hbox{$\, \buildrel {\scriptstyle <}\over {\scriptstyle\sim}\,$}}

\newcommand{\chandra}{{\emph{Chandra}}}

\newcommand{\ie}{i.e.,\,}
\newcommand{\be}{\begin{equation}}
\newcommand{\ee}{\end{equation}}
\newcommand{\bea}{\begin{eqnarray}}
\newcommand{\eea}{\end{eqnarray}}
\newcommand{\nhat}{\hat{\bf n}}

\shorttitle{\emph{CHANDRA} OBSERVATIONS of GRAVITATIONAL LENSES}
\shortauthors{CHEN ET AL.}

\tighten

\begin{document}

\def\sarc{$^{\prime\prime}\!\!.$}
\def\arcsec{$^{\prime\prime}$}

\title{Effects of Kerr Strong Gravity on Quasar X-ray Microlensing }

\author{Bin Chen\altaffilmark{1}, Xinyu Dai\altaffilmark{1}, E. Baron\altaffilmark{1,2,3}, R. Kantowski\altaffilmark{1}}

\altaffiltext{1}{Homer L. Dodge Department of Physics and Astronomy, The University of Oklahoma,
Norman, OK, 73019, USA, bchen@ou.edu}
\altaffiltext{2}{Computational Research Division, Lawrence Berkeley
        National Laboratory, MS 50F-1650, 1 Cyclotron Rd, Berkeley, CA}
\altaffiltext{3}{Hamburger Sternwarte, Gojenbergsweg 112, 21029 Hamburg, Germany }

\begin{abstract}
Recent quasar microlensing observations have constrained the sizes of X-ray emission regions to be within about $10$ gravitational radii of the central supermassive black hole.
Therefore, the X-ray emission from lensed quasars is first strongly lensed by the black hole before it is lensed by the foreground galaxy and star fields.
We present a scheme that combines the initial strong lensing of a Kerr black hole with standard linearized microlensing by intervening stars.
We find that X-ray  microlensed light curves incorporating Kerr strong gravity can differ significantly from standard  curves.
The amplitude of the fluctuations in the light curves can increase or decrease by $\sim$0.65--0.75 mag by including Kerr strong gravity.
Larger inclination angles give larger amplitude fluctuations in the microlensing light curves.
Consequently, current X-ray microlensing observations might have under or overestimated the sizes of the X-ray emission regions.  
We estimate this bias using a simple metric based on the amplitude of magnitude fluctuations. 
The half light radius of the X-ray emission region can be underestimated up to $\sim$50\% or overestimated up to $\sim$20\%.
Underestimates are found in most situations we have investigated.
The only exception is for a  disk with large spin, radially flat emission profile, and observed nearly face on, where an overestimate is found.    
Thus, more accurate microlensing size constraints should be obtainable by including Kerr lensing.
The caustic crossing time can differ by months after including Kerr strong gravity.  
A simultaneous monitoring of gravitational lensed quasars in both X-ray and optical bands with densely sampled X-ray light curves might reveal this feature.
We  conclude that it  should be possible to constrain important parameters such as inclination angles and black hole spins from combined Kerr and microlensing effects.
\end{abstract}


\section{Introduction}\label{sec:intro}

It is well-known that most active galactic nuclei (AGN) are copious X-ray emitters.
However, despite efforts for decades, the origin of this emission is still unclear.
Standard AGN accretion disk theory predicts disk  temperatures ($\rm \lesssim 10^5\,K $) too low to emit X-rays ($\sim$$\rm 10^9\,K$ for hard X-rays).
Instead, the X-ray emission is generally believed to be generated by reprocessing of optical/UV disk photons by hot electrons in a corona above the accretion disk via unsaturated multiple inverse Compton scattering.
The two-phase (cold disk plus hot corona) accretion disk model (Haardt \& Maraschi 1991, 1993) predicts that the observed spectrum contains three components: a direct power-law component from the X-ray corona (Zdziarski et al.\ 1994), a reflection component from the disk (Guilbert \& Rees 1988; Lightman \& White 1988) with metal emission lines (in particular, the \feka\ line, Fabian 1989; Laor \& Netzer 1989), and thermal radiation from the cold accretion disk (Shakura \& Sunyaev 1973).
Despite the general belief in the existence of the X-ray corona, fundamental questions remain unanswered: what is its size and geometrical structure, how are its hot electrons heated, and why does it radiate so much ($\sim$10\%) of the accretion luminosity. 

A major difficulty in solving the AGN corona puzzle is that its small angular size makes it unresolvable by current telescopes.
Many estimates of the corona size are based on variability arguments.
The observed rapid X-ray variability (as short as a few hours) was important early evidence for small X-ray emission  sizes.
Recently techniques based on a Bayesian Monte-Carlo analysis method (Kochanek 2004; Poindexter \& Kochanek 2010) together with accumulated high quality data (e.g., Chen et al.\ 2012a) are making quasar microlensing a very powerful tool in constraining corona geometry (Blackburne et al.\ 2006;  Pooley et al.\ 2006; Morgan et al.\ 2008; Chartas et al.\ 2009; Dai et al.\ 2010; Blackburne et al.\ 2011, 2012; Morgan et al.\ 2012).
These observations have conclusively constrained the quasar X-ray emission size to be of order $\sim$$10\,r_g$ ($r_g\equiv\rm GM_{\rm BH}/c^2$, the gravitational radius), much smaller than the optical emission size.
Furthermore, Chen et al.\ (2011) detected energy-dependent X-ray microlensing in Q~2237+0305, and their results suggest that the hard X-ray emission might come from regions smaller than the soft X-rays.
Based on \chandra\ monitoring data for six gravitationally lensed quasars,  Chen et al.\ (2012a) found that the rest frame equivalent widths of the \feka\ line are significantly higher than those measured in typical AGN.
This suggests that the iron line emitting region is more compact than the continuum emission region.
 If true, studying quasar X-ray emission (continuum or metal lines) probes the innermost regions of AGN, the region where relativistic effects of the central black hole are important.

Quasar X-ray microlensing observations study the gravitational lensing of the X-ray emission by  random foreground star fields in intervening galaxies.
However, the interpretation of these observations assumes a flat space-time for the source plane, and constrains the {\it projected} source area along the line of sight.
When the X-ray source is within a few gravitational radii of the central supermassive black hole  powering the AGN, both the flux profile and images of the X-ray emission are significantly altered (Bardeen et al.\ 1972; Cunningham 1975; Fabian et al.\ 1989;  Chen et al.\ 1989; Laor et al.\ 1989; Laor 1991; Rauch \& Blandford 1994; Bromley et al.\ 1997;  Beckwith \& Done 2004; Popovi$\rm\acute{c}$ et al.\ 2006; Schnittman \& Krolik 2010; Abolmasov \& Shakura 2012; Chen et al.\ 2012b).
In other words,  a quasar's X-ray emission is gravitationally lensed by the central black hole at the very beginning of  its trip to a distant observer,  well before it is lensed by any foreground mass concentrations.
We refer to the strong lensing by a Kerr black hole as ``Kerr lensing" throughout this paper.
Kerr lensing differs from standard linear lensing theory in a few important aspects (Chen et al.\ 2012b), and we emphasize two of them here.
First, the linear approximation is not valid since the X-ray sources are very close to the central black holes (a few $r_g$),  and consequently, there is no simple lens equation which can be used to find the source position given the image position on the sky.
Second, the redshifts produced by the gravity field of the central black hole and the relativistic motion of the X-ray source are very important for Kerr lensing in contrast to standard lensing theory in which the redshift comes solely from cosmology.
Strong bending and redshift effects of Kerr lensing change the shape, size, and profile of the X-ray source,  all of which are input parameters for the foreground microlensing.
The constraints obtained from current quasar X-ray microlensing  should consequently be expanded by including Kerr gravity.
 Because Kerr lensing depends on important parameters such as inclination angle, black hole spin, and the velocity flow of the X-ray sources, quasar microlensing might become an even more powerful tool in probing the innermost region of AGN.
We investigate these ideas in this paper.

\section{Combining Kerr Lensing and Microlensing}

The space-time of a black hole with nonzero angular momentum is described by  the Kerr metric (Kerr 1963) which depends only on two parameters: the black hole mass $M_{\rm BH}$ and spin $a$.
As stated earlier, there is no simple lens equation for Kerr strong lensing.
We have developed a ray-tracing code using a $5^{\rm th}$ order Runge-Kutta algorithm with adaptive step size control in Chen et al.\ (2012b) to remedy this problem (see also Dexter \& Agol 2009, and Vincent et al\ 2011).
Ray-tracing has also become a standard technique in gravitational microlensing (e.g., Kayser et al.\ 1986; Schneider \& Weiss 1987).
In standard lensing theory photons travel along straight lines, except at the lens plane where bending happens instantaneously.
The time consuming part in generating microlensing magnification patterns in the source plane is adding the bending angles caused by all the random stars (from a few tens to tens of thousands depending on the specific problem).
To find the magnification in pixels near the caustics where the magnification diverges for point sources, thousands of (or even more) rays need to be traced.
Algorithms based on a hierarchical tree code (Wambsganss 1999),  Fourier transforms (Kochanek 2004) or tessellation method (Mediavilla et al.\ 2006) have been developed to improve the speed of this process.
In contrast to ray-tracing in microlensing where the bending angle needs to be computed only once for each ray, for Kerr lensing, it can take a few hundred steps to trace a ray from a distant observer (e.g., $r_{\rm obs}=10^6\, r_g$) to a point  on the accretion disk near the black hole with desired relative accuracy, e.g., $10^{-9}$. 
Smaller and smaller  step sizes are used in our code to obtain the desired accuracy when the photon is close to the black hole, and for each step, many computations involving the (complicated) Kerr metric tensor must be made.
We combine Kerr lensing and microlensing in  the following.

\begin{figure}
	\epsscale{0.8}
	\plotone{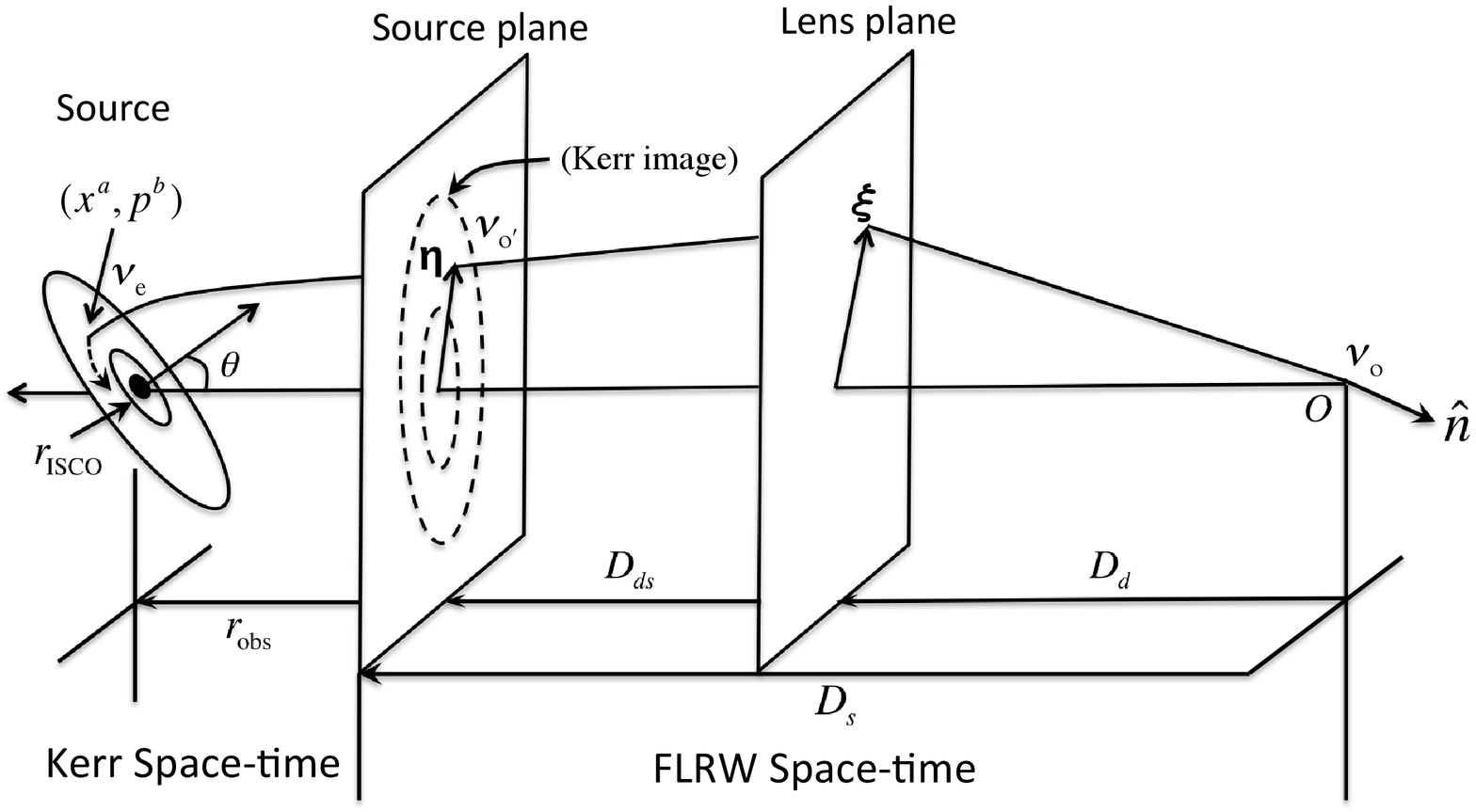}
	\caption{Schematic light ray path for Kerr+Micro lensing raytracing.
	Here $\theta$ is the disk inclination angle, and $r_{\rm ISCO}$ is the innermost stable circular orbit.
	A ray arriving at the observer in a direction $\hat{\bf n}$ is backward traced to $\mbox{\boldmath$\xi$}$ (lens plane), then to $\mbox{\boldmath$\eta$}$ (source plane) by microlensing raytracing, and then to $(x^a,p_a)$ on the X-ray source near the Kerr black hole by Kerr ray-tracing.
	 The photon frequencies $\nu_{\rm o},$ $\nu_{\rm o'},$ and $\nu_{\rm e}$ are measured at the observer, source plane, and the source, respectively.
	The Kerr image (in the source plane) is treated as the source for the foreground microlensing ray-tracing.
	$r_{\rm obs}$ is greatly exaggerated  with respect to the angular diameter distance $D_s.$
	 \label{fig:ray-tracing}}
\end{figure}

We choose $M_{\rm BH}= 10^9M_{\odot}$ as the mass of the supermassive black hole (e.g.\ Bian \& Zhao 2002; Wang et al.\ 2003) and assume the lens and source redshifts $z_d=0.5$ and $z_s=1.0,$ respectively.
We assume  a flat Friedman-Lema\^itre-Robertson-Walker (FLRW) cosmology with $\Omega_{\rm m}=0.3$ and $\Omega_\Lambda=0.7.$
The angular diameter distance of the source is $D_s=1653\,{\rm Mpc}=3.45\times10^{13}\, r_g.$
We choose the source plane for the foreground microlensing at the distance $r_{\rm obs} = 10^6\, r_g$ from the Kerr black hole toward the observer and orthogonal to the line of sight (Figure~\ref{fig:ray-tracing}).
Because the most significant non-vanishing components of the Riemann curvature tensor of the Kerr space-time are of order $M_{\rm BH}/r^3,$ we are safe in ignoring the curvature from the Kerr black hole at this distance and  in assuming this plane is flat for the foreground microlensing raytracing.
On the other hand, since $r_{\rm obs}/D_s$ is of order $10^{-7},$ the microlensing magnification pattern in this plane is numerically the same as a (fictitious) flat plane at redshift $z_s.$
In short, the backward raytracing for the foreground gravitational microlensing can be done with no other modifications. 

If we choose $m_{\rm star}= 0.3 M_{\odot}$ as a representative stellar mass for the foreground microlensing, the Einstein ring angle is $\theta_E\approx 4.4\times 10^{-12}\,\rm rad$ (or $D_s\theta_E =152.6\, r_g$). 
Suppose we want to generate a microlensing magnification pattern in the source plane (Figure~\ref{fig:ray-tracing}) for a window of size $10\,\theta_E\times 10\,\theta_E$ with 1024 pixels in each dimension, the pixel size is $\sim$$1.5\,r_g.$
On the other hand, if we start the backward ray-tracing for Kerr lensing from the source plane ($10^6\,r_g$ from the black hole), using $\theta_E$ as the characteristic bending angle of foreground microlensing, the error induced by treating rays starting from the source plane as parallel light toward the black hole  is of order $r_{\rm obs}\theta_E=10^{-6}r_g,$ much smaller than the resolution needed in the foreground microlensing even if we increase the desired resolution by factor of 1000 to a pixel size of $\sim$$0.001\,r_g.$
If the mass of the central massive black hole is $10^{6} M_{\odot}$ or $10^{12}M_{\odot}$ instead of $10^9M_{\odot}$ as used in this paper, the parallel light approximation for the Kerr lensing will still be valid.

To compute the flux at the observer from an extended source near the black hole which is lensed by both the supermassive black hole and the foreground microlensing stars, we need two quantities.
The first is the magnification pattern in the source plane caused by foreground microlensing (Figure~\ref{fig:ray-tracing}).
This can be generated independently of the Kerr lensing, although the required resolution depends on the source size.
The second quantity we need is the Kerr lensing image of the source on the source plane including its intensity profile.
This can be computed using our Kerr ray-tracing code with the parallel light approximation, independently of  the foreground microlensing.
For each pixel in the source plane we need only trace one ray,  e.g., from the center of this pixel toward the black hole, instead of hundreds of rays as in microlensing.
This greatly saves time used for Kerr ray-tracing.
In short, an extended light source near the black hole is (inversely) mapped onto the nearby source plane (placed orthogonal to the line of sight and at rest in the background cosmology) by using the Kerr ray-tracing code.
This mapping includes the image, its redshift, and its intensity profile. Kerr+Micro lensing light curves are then generated by using the  2-D Kerr image as the  source for the microlensing magnification pattern.

 Let $\nu_{\rm o}$ be the photon's frequency  measured by a cosmic observer when the photon is received and  $\nu_{\rm e}$ the frequency measured by an observer at rest with respect to the corona at emission.
We split the redshift into two parts,  the cosmological redshift, and  the ``Kerr redshift"
\be
\frac{\nu_{\rm o}}{\nu_{\rm e}}=\frac{\nu_{\rm o}}{\nu_{\rm o'}}\frac{\nu_{\rm o'}}{\nu_{\rm e}}=\frac{1}{1+z_s}\frac{\nu_{\rm o'}}{\nu_{\rm e}}=\frac{g}{1+z_s}
\ee where $\nu_{\rm o'}$ is the photon frequency measured in the source plane defined above.
The Kerr redshift factor $g\equiv {\nu_{\rm o'}}/{\nu_{\rm e}}$ depends on the inclination angle $\theta$ of the accretion disk, the 2-D impact vector $\mbox{\boldmath$\eta$}$ in the source plane with respect to the optical axis, and on the relativistic flow of the corona (Figure~\ref{fig:ray-tracing}).
Let $I(x^a,p^b)$ be the specific intensity profile of the source (e.g., the hot corona emitting X-rays) where $(x^a,p^b)$ are the 7-D phase space coordinates ($p^ap_a=0$ for massless particles), the observed monochromatic flux is
\bea\label{flux_general}
F_{\nu_{\rm o}}&=&\int_{}{I^{\rm obs}_{\nu_{\rm o}}(\nhat)\cos\vartheta d\Omega_{\rm o}}
=\int_{}{I^{\rm obs}_{\nu_{\rm o}}(\nhat)d\Omega_{\rm o}}
=\int_{{\cal D}_L}{\frac{I_{(1+z_d)\nu_{\rm o}}(\mbox{\boldmath$\xi$})}{(1+z_d)^3}\frac{d^2\xi}{D_d^2}  }
=\int_{{\cal D}_S}{\frac{I_{\nu_{\rm o'}}(\mbox{\boldmath$\eta$})}{(1+z_s)^3}{\cal A}(\mbox{\boldmath$\eta$})\frac{d^2\eta}{D_s^2}}\cr
&=&\frac{1}{(1+z_s)^3}\frac{1}{D_s^2}\int_{{\cal D}_S}{g^3(\mbox{\boldmath$\eta$})I[x^a(\mbox{\boldmath$\eta$}), p^a(\mbox{\boldmath$\eta$})]{\cal A}(\mbox{\boldmath$\eta$})d^2\eta}
\eea
where ${\cal D}_L$ and ${\cal D}_S$ are 2-D integral domains (i.e., the images of the X-ray source) in the foreground lens plane and the source plane, respectively, ${\cal A}$ is the microlensing magnification in the source plane, and the composite map $\mbox{\boldmath$\xi$}\rightarrow\mbox{\boldmath$\eta$}\rightarrow (x^a,p^a)$  is realized by microlensing and Kerr raytracing, respectively (we have used $\cos\vartheta \approx 1$ in the second step).
The effect of foreground microlensing is contained in the magnification factor ${\cal A}$.
The effect of Kerr lensing is multifold:  a) the integral domain ${\cal D}_S$ for the case of Kerr lensing is enlarged and distorted compared with that of the case without Kerr lensing; b) the intensity profile of the Kerr image is changed by the Kerr light bending, e.g., a pixel in the source plane will be mapped to  a different place in the corona from that of straight line ray-tracing; c) the Kerr redshift  combined with relativistic beaming also changes the profile significantly (e.g., the $g^3$ factor in the integral above).

\section{Examples and Results}

As an example, we assume a simple double power-law in radius and frequency model to the X-ray emission,
\be\label{X-ray_profile}
I_\nu(\nu, r) \propto \frac{1}{r^n}\frac{1}{\nu^{\Gamma-1}},
\ee where $r$ is the radial coordinate of the Kerr metric, $n$ specifies the steepness of the radial profile, and $\Gamma$ is the photon index.
Using Eq.~(\ref{flux_general}) the observed monochromatic flux at frequency $\nu_{\rm o}$ is
\bea\label{F_nu}
F_{\nu_{\rm o}}=\frac{1}{(1+z_s)^{\Gamma+2}}\frac{1}{D_s^2}\frac{1}{\nu_{\rm o}^{\Gamma-1}}\int_{{\cal D}_S}{\frac{g^{\Gamma+2}(\mbox{\boldmath$\eta$})}{r^n(\mbox{\boldmath$\eta$})}{\cal A}(\mbox{\boldmath$\eta$})d^2\eta}
\eea (we have dropped the unimportant constant).
We take the customary $\Gamma=2.0$ used for quasar X-ray emission (Chen et al.\ 2012a), and $n=3$ or $0$ (the source profile is radially steeper, and more concentrated toward the center for $n=3$).
We choose a simple geometry for the X-ray source:  a thin X-ray disk immediately above the accretion disk moving with Keplerian flow.
This is partially motivated by the ``sandwich" corona model (Haardt \& Maraschi 1991, 1993).
We limit the radial extent of the emission to $r_{\rm ISCO}<r< r_{\rm disk}$ where $r_{\rm inner}=r_{\rm ISCO}$ (the innermost stable circular orbit, Bardeen et al.\ 1972) and $r_{\rm disk}\le 50\,r_g.$ 
We focus on the $r_{\rm disk} = 20\,r_g$ case in our analysis.
An X-ray emission region of this size is consistent with existing quasar microlensing observations (see Section~\ref{sec:intro}).
Since most of the current constraints from quasar X-ray microlening are upper bounds, a smaller size is possible (Morgan et al.\ 2012).
The Kerr lensing effect will be more important for smaller  X-ray emitting region. 
As for the spin of the black hole, we experimented with $a=0$ (a Schwarzschild black hole) and $a=0.998M_{\rm BH}$ (an extreme Kerr black hole, Thorne 1974).
The inner cutoff $r_{\rm ISCO}$ depends on the black hole spin and is $1.24\,r_g$ and $6\,r_g$ for $a=0.998M_{\rm BH}$ and $0$, respectively.
The minimum $r_{\rm disk}$ we considered is $2.5\,r_g$ and $10\,r_g$ for $a=0.998M_{\rm BH}$ and $a=0,$ respectively.
We compare results for three inclination angles $\theta=15^\circ,$ $45^\circ,$ and $75^\circ,$ covering the range from nearly face on to nearly edge on.
Figure~\ref{fig:intensity_image} shows intensity images of lensed X-ray emitting disks for $r_{\rm disk}=50\,r_g.$ 
Figure~\ref{fig:hl_radius} and Table~\ref{tab:hl_radius} show the strong lensing corrections to half light radii of X-ray disks (see Section~\ref{sec:hl_radius}).

\begin{figure*}
\begin{center}$
\begin{array}{cc}
\includegraphics[width=0.55\textwidth,height=0.42\textheight]{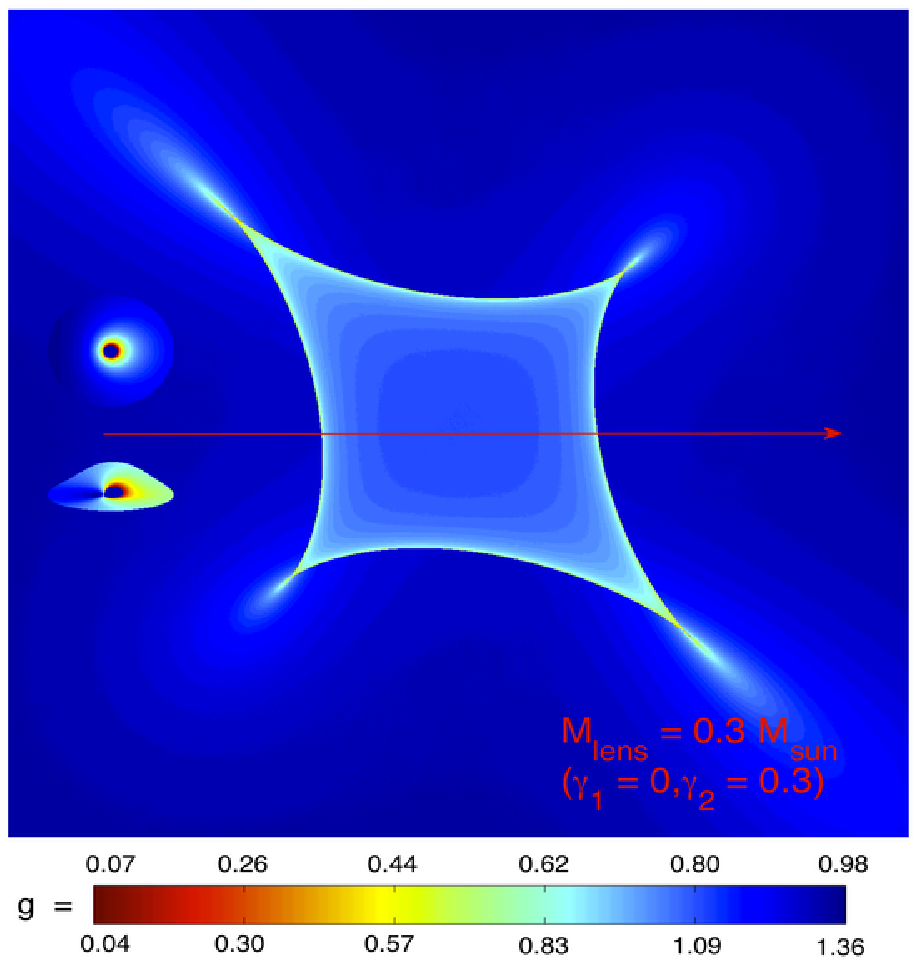}
\hspace{0pt}
\includegraphics[width=0.55\textwidth,height=0.42\textheight]{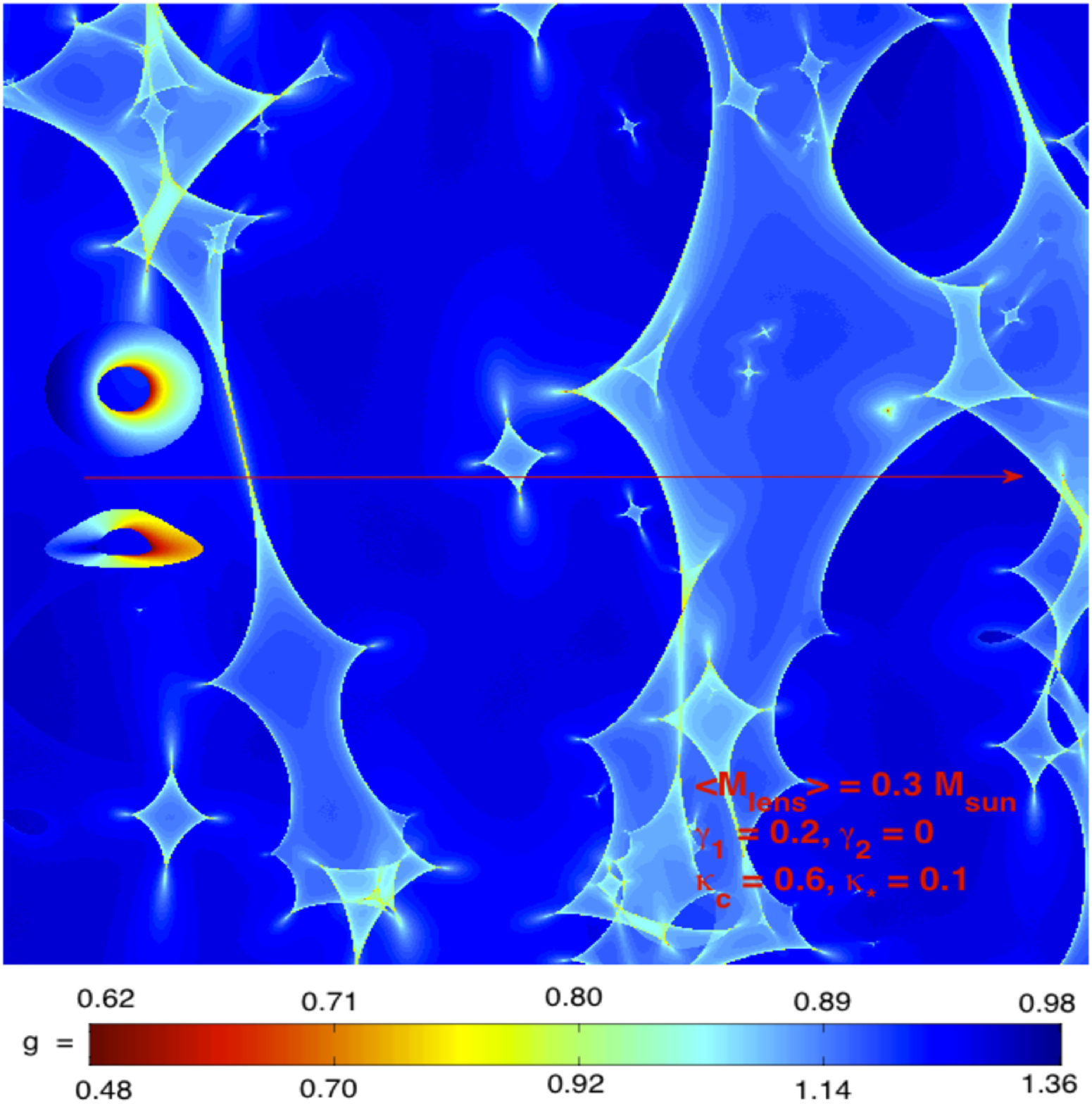}
\end{array}$
\end{center}
\caption{Magnification pattern (in the source plane) of a simple Chang-Refsdal lens ($2\,\theta_E\times 2\,\theta_E$) and a random star field ($9\,\theta_E\times 9\,\theta_E$).
	The resolutions are $6000\times 6000$ (left) and $7200\times 7200$ (right).
	The insets are the Kerr lensing images of an X-ray disk of size $r_{\rm disk}=20\, r_g$ by a supermassive black hole ($M_{\rm BH}=5\times10^8M_{\odot}$ left, $10^9 M_{\odot}$ right) with spin $a=0.998M_{\rm BH}$ (left panel) and $a=0$ (right panel)  observed at inclination angle $\theta=15^\circ$  and $75^\circ$ (upper/lower insets), respectively.    	
	We scaled the disk image by a factor of 2 (left panel) and 5 (right panel) with respect to the source window for clarity.
	The thin red lines  are trajectories of the X-ray source in the source plane.
  	The color-bars show the Kerr redshift factor $g=\nu_{\rm o'}/\nu_{\rm e}$ for $\theta=15^\circ$ (first row) and $\theta=75^\circ$ (second row).
	A disk with higher inclination angle, or larger spin (smaller $r_{\rm ISCO}$) spans a larger redshift interval.
	\label{fig:mag_pattern}}
\end{figure*}

We considered two foreground microlensing models:  a simple Chang-Refsdal lens (Chang \& Refsdal 1979) with $M_{\rm lens}=0.3M_\odot,$ and external shear $\gamma=0.3,$ and a random star field with mean lens mass $\langle M_{\rm star}\rangle=0.3M_{\odot}$ ($0.01M_{\odot}\le M_{\rm star}\le 1.6M_{\odot}$), $\kappa_*=0.1$, $\kappa_c=0.6,$ and shear $\gamma=0.2$ (Schneider et al.\ 1992).
The simple, but nontrivial caustic structure of a Chang-Refsdal lens is ideal for studying the effect of Kerr lensing  on foreground microlensing.
The random star field model is more realistic for a quasar at cosmic redshift seen through a foreground lensing galaxy.
The magnification patterns for the two models are shown in Figure~\ref{fig:mag_pattern}.
We also show the redshift images of lensed X-ray disks in the source plane for inclination angles of $\theta=15^\circ$ and $75^\circ,$ and black hole spins of $a=0.998M_{\rm BH}$ and $0.$ 
The microlensing light curves for the source trajectories  shown in Figure~\ref{fig:mag_pattern} (red lines) are shown in Figures~\ref{fig:LC_Chang} and \ref{fig:LC_shear} for $r_{\rm disk}=20\,r_g$ where we compare Kerr+Micro lensing with microlensing variability.
We show the magnitude $m\equiv 2.5\log_{10}\mu,$ where $\mu\equiv F^{\rm lensed}_{\nu_{\rm o}}(\theta)/F^{\rm unlensed}_{\nu_{\rm o}}(\theta)$ is the lensing magnification.
Using this ratio modulates out the source luminosity and  the $\cos\theta$ part of the projection effect.
We show the light curves for four source models ($a=0.998M_{\rm BH}$ or 0, $n=3$ or 0), and three inclination angles $\theta = 15^\circ,$ $45^\circ,$ and $75^\circ$ in the first and third rows.
We also show the inclination dependence of the Kerr+Micro lensing light curves by plotting $m(t;45^{\rm o})-m(t;15^{\rm o})$ and $m(t;75^{\rm o})-m(t;15^{\rm o})$ in the second and fourth rows (dashed and dotted curves respectively) with and without the projection effect (\ie the $\cos\theta$ factor, Cyan and Magenta respectively).
The spin dependence of the Kerr+Micro lensing is shown by $m(t;a=0.998M_{\rm BH})-m(t;a=0)$ in the fifth rows.
One of the important quantities of a microlensing light curve is the amplitude of the magnitude fluctuation, \ie the difference between the maximum and minimum magnitude, $m_{\rm max}-m_{\rm min}\equiv m_{\rm max-min}.$
This quantity can be used to estimate the source size because light curves of smaller sources have larger amplitudes of fluctuations. 
We tabulate this quantity for the light curves in Figures~\ref{fig:LC_Chang} and \ref{fig:LC_shear} in Table~\ref{tab:dMag} ($r_{\rm disk}$ is fixed to be $20\,r_g$).
We furthermore compute this quantity for a range of source sizes $2.5\,r_g < r_{\rm disk}<50\,r_g$ restricted to the random star field model (the right panel of Figure~\ref{fig:mag_pattern}).
We show the results in Figure~\ref{fig:max_min_metric} and Table~\ref{tab:delta_m_metric}. 

\subsection{Inclination Angle and Spin Dependence of Kerr+Micro Lensing Light Curves}

\begin{figure}
	\epsscale{0.75}
	\plotone{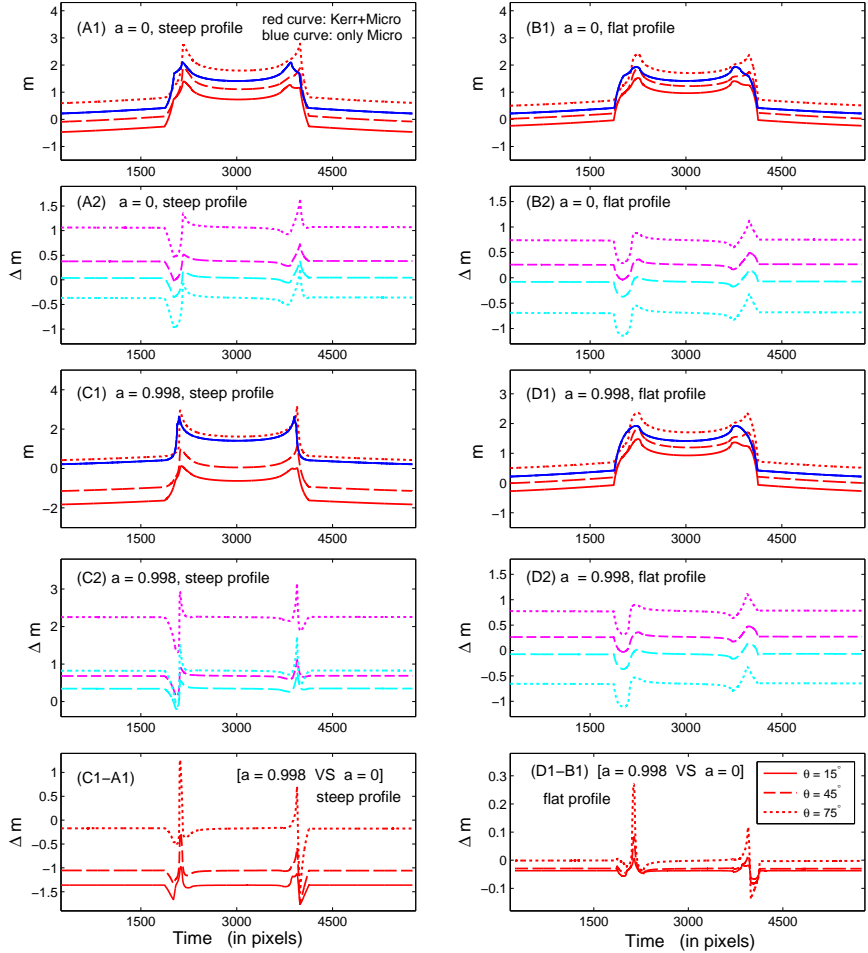}
	\caption{X-ray microlensing light curves for the Chang-Refsdal lens model, showing the magnification in magnitudes ($m\equiv2.5\log_{10}{\mu},$ and $\mu$ is the lensing magnification) as a function of time in pixel units.
	We show results for 4 source models,  spin $a = 0$  or  $0.998 M_{\rm BH}$, and radial profile $n=3$ (steep, the first column) or 0 (flat, the second column) as indicated in the panels.
	The solid,  dashed, and dotted curves are for inclination angle $\theta=15^\circ$, $45^\circ,$ and $75^\circ,$ respectively.
	The red and blue curves are respectively for Kerr+Micro lensing or microlensing only.
	The dependence of standard microlensing curves on the inclination angle is so weak that it is hard to distinguish these curves by eye.
	In the second and the fourth row, we show the inclination dependence of the Kerr+Micro lensing curves by plotting $m(t;45^{\rm o})-m(t;15^{\rm o})$ and $m(t;75^{\rm o})-m(t;15^{\rm o})$ (magenta dashed and dotted curves respectively). 
	The Cyan curves plot the same things but including the geometrical projection effect (the $\cos\theta$ factor).  
	In the fifth row, we show the spin dependence of the Kerr+Micro lensing light curves by plotting $m(t;a=0.998)-m(t;a=0)$ for each inclination angle.
		 \label{fig:LC_Chang}}
\end{figure}

\begin{figure}
	\epsscale{0.9}
	\plotone{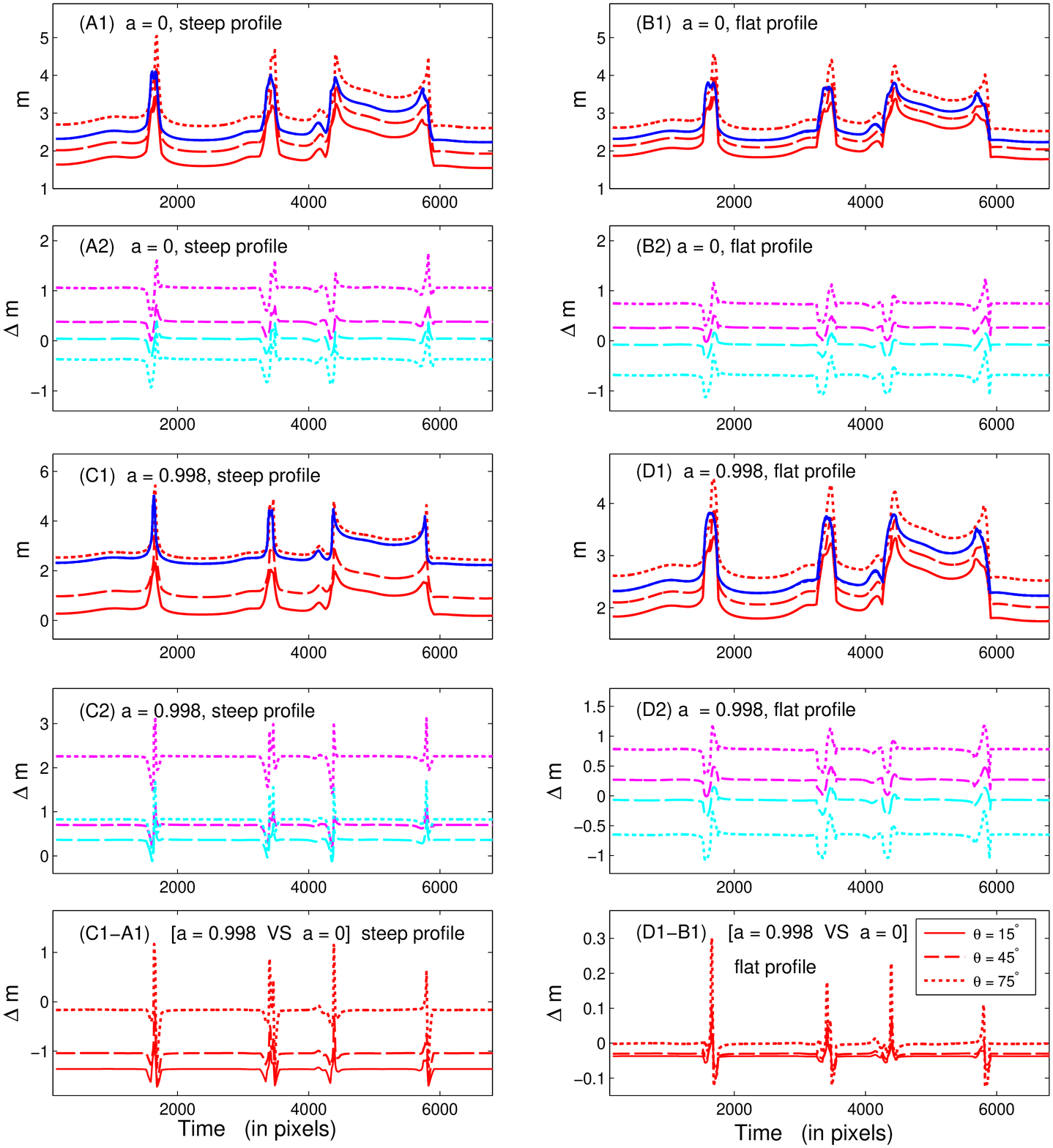}
	\caption{X-ray microlensing light curves for the random star field lens model with $\langle M_{\rm lens}\rangle=0.3 M_{\odot}$, $\kappa_c=0.6$ (continuous mass),  $\kappa_*=0.1$ (stellar mass), and shear $\gamma_1=0.2$, $\gamma_2=0$.
	The definition of each line is the same as Figure~\ref{fig:LC_Chang}.
	 \label{fig:LC_shear}}
\end{figure}

The effects of Kerr strong gravity on microlensing can be seen from the light curves of both foreground lens models (Figures~\ref{fig:LC_Chang} and \ref{fig:LC_shear}).
The most important result is that the Kerr+Micro lensing light curve depends significantly on the inclination angle, while the standard microlensing light curve does not. 
Beyond the $\cos\theta$ projection effect, which we have factored out, the standard microlensing light curves for three inclination angles, shown in blue curves, are almost indistinguishable from each other, see Figures~\ref{fig:LC_Chang}, \ref{fig:LC_shear}, and Table~\ref{tab:dMag}. 
For small inclination angles the Doppler shift is unimportant, whereas the gravitational redshift significantly reduces the flux of the X-ray source.
For high inclination angles, the Doppler effect becomes more important as the flux of a source moving with relativistic speed is strongly magnified or demagnified depending on whether it is approaching or receding from the observer (relativistic beaming).
We find that the intensity profile of the Kerr image in the source plane can be significantly different from that of the source.
The effect of image distortion is also more significant for high inclination angles (see Figure~\ref{fig:intensity_image}).
In regions of the source plane without complicated caustic structures the gradient of the magnification is small, and the  Kerr+Micro lensing light curves differ from that of microlensing by roughly a constant, \ie by the amount the flux at the source plane is reduced (or  increased for high inclination angle cases) by Kerr lensing.
However, when the source is crossing a caustic or regions with clustered caustic structure (see Figure~\ref{fig:mag_pattern}), the difference between  the two lensing schemes deviates from merely a constant (see Figures~\ref{fig:LC_Chang} and \ref{fig:LC_shear}).
Kerr lensing changes the size, shape, and profile of the source which are inputs for the foreground microlensing.
This can also be seen from Table~\ref{tab:dMag} for both the Chang-Refsdal lens model and the random star field model. 
 For $r_{\rm disk}=20\,r_g$, the amplitude of the magnitude fluctuation, $m_{\rm max-min}$ can differ by $\sim$0.5--0.6 mag  between a flat X-ray disk and a Kerr-lensed disk for both foreground lens models considered in this paper. 
This deviation from a constant shift is inclination angle dependent, and a larger inclination angle results in a larger amplitude of magnitude fluctuation,  as summarized in Table~\ref{tab:dMag} (\ie $m_{\rm max-min}$ increases with $\theta$ for Kerr+Micro lensing).
For a corona with a steep radial intensity profile ($n=3$) observed nearly face on, the more severe gravitational redshift suffered by the central region of the corona makes the source emission less concentrated toward the center, and consequently, the Kerr+Micro lensing light curves show smaller fluctuation amplitudes than standard microlensing light curves.
For an observer at high inclination angle, the X-ray emission is more concentrated in Doppler blue-shifted regions by relativistic beaming, and consequently, we see larger fluctuation in the Kerr+Micro lensing light curves.  
The influence of Kerr lensing on X-ray microlensing light curves is more significant for source emission with steeper radial profiles or larger spins (this latter gives smaller  $r_{\rm ISCO}$ values) both of which result in more concentrated X-ray emission in regions close to the event horizon where the effect of the black hole is most important.

In Figure~\ref{fig:max_min_metric} we show $m_{\rm max - min}$ as a function of emission size $r_{\rm disk}$ for four background source models (a = 0 or $0998M_{\rm BH},$ $n= 0$ or 3) and three inclination angles restricted to the random star field model.
The data is given in Table~\ref{tab:delta_m_metric}.
The inclination angle dependence of $m_{\rm max - min}$ is much more significant for Kerr-lensed disks than for flat disks (the results for flat disks, shown as dashed curves, are almost indistinguishable from each other). 
Larger inclination angles give larger amplitude fluctuation in the microlensing light curves.
The amplitude of magnitude fluctuation, $m_{\rm max-min},$ can increase by $\sim$0.65 mag or decrease by $\sim$0.75 mag including Kerr strong gravity (refer to the (1,1) and (2,1) panels of Figure~\ref{fig:max_min_metric}).
If a very large amplitude of magnitude fluctuation is observed, e.g., $m_{\rm max-min}\gtrsim 3.2,$ then a Kerr-lensed disk is strongly favored (none of the four flat models can produce $m_{\rm max-min}$ of this size, see Table~\ref{tab:delta_m_metric}). 
An even larger amplitude of fluctuation will calls for a steeper radial profile ($n>3$), extra X-ray emission within the $r_{\rm ISCO},$ or very large inclination angles (to focus the emission by relativistic beaming). 
Since the amplitude of magnitude fluctuation, $m_{\rm max-min},$ of a Kerr-lensed disk can be either larger or smaller than that of a flat disk, current quasar X-ray microlensing observations might have  over or underestimated the X-ray emission sizes by ignoring Kerr strong gravity.
We will discuss this point in detail in \textsection\ref{sec:hl_radius}.

The spin dependence of microlensing light curves (with or without Kerr strong lensing) is shown in the third row of Figure~\ref{fig:max_min_metric}.
We show the difference in $m_{\rm max-min}$, \ie $\Delta m_{\rm max-min}\equiv m_{\rm max-min}^{(a=0.998)}-m_{\rm max-min}^{(a=0)}$ as a function of $r_{\rm disk}$ for two radial profiles ($n=0$ or 3) and for three inclination angles.
For the case of a steep radial profile, the spin dependence of $m_{\rm max-min}$ of a flat X-ray disk is in fact more significant than a Kerr-lensed disk, mainly because an $a=0.998M_{\rm BH}$ disk has a smaller $r_{\rm ISCO}$ ($1.24\,r_g$) than an $a=0$ disk ($r_{\rm ISCO}=6\,r_g$) and therefore has a more concentrated emission profile (or smaller half light radius, see \textsection\ref{sec:hl_radius}).
Consequently, the $a=0.998M_{\rm BH}$ case shows a much larger $m_{\rm max- min}$ than $a=0$ case.\footnote{This  spin dependence of microlensing light curves for flat X-ray disks is physically simple, but does not seem like have been pointed out previously.}
For Kerr-lensed disks,  the concentrated emission in the region $1.24\,r_g<r<6\,r_g$ was washed out by gravitational redshifts, in particular, for the nearly face on case ($\theta=15^\circ$), see the $(3, 1)$ panel of Figure~\ref{fig:max_min_metric}.
For a flat radial profile, the un-lensed source emission is uniformly distributed over the X-ray disk, therefore, the extra emission between $1.24\,r_g<r<6\,r_g$ does not contribute significantly to the total flux. 
Consequently, the spin dependence of both flat and Kerr-lensed disks is insignificant.

\subsection{Strong Lensing Correction to  Half light Radii of X-ray Coronae}\label{sec:hl_radius}

\begin{figure}
	\epsscale{1.0}   
	\plotone{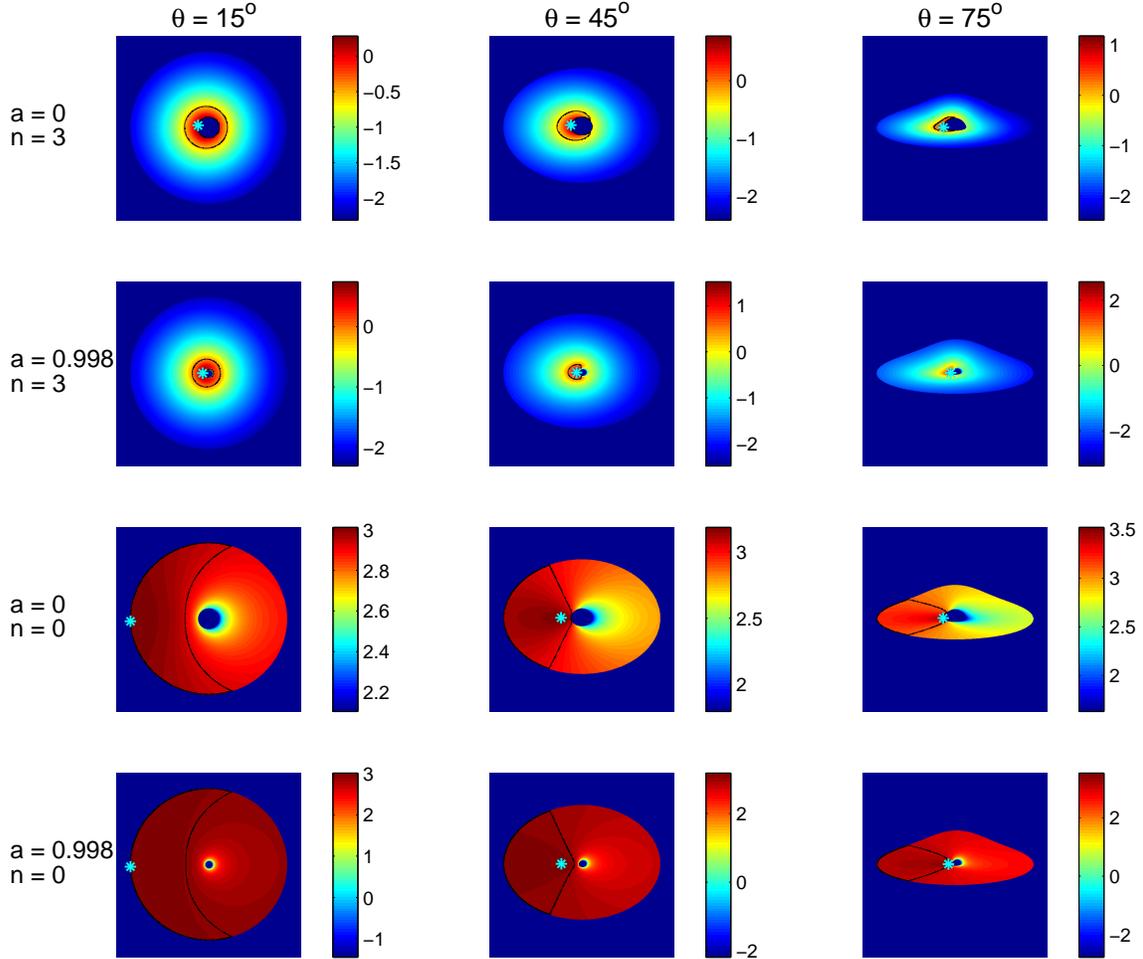}
	\caption{Intensity plots of Kerr lensed X-ray emitting disks moving with Keplerian flow. 
	The color-bars are in logarithmic scale (normalization is arbitrary).
	$r_{\rm disk}= 50\, r_g.$ 
	We show results for 4 source models (spin $a =0$ or $0.998\,M_{\rm BH}$ and radial profile $n=3$ or $0$) in the 4 rows, and three inclination angles $\theta = 15^{\rm o}$, $45^{\rm o}$ and $75^{\rm o}$ in the first, second and third column.  
	The star in each panel marks the peak surface brightness location. 
	For nearly face on cases with flat radial profile ($n=0$), the peak brightness happens on the left boundary of the disk (the approaching side) because of gravitational and Doppler shifts. 
	The black curve in each panel is the critical intensity contour line delimiting the half light region (used to compute the half light radius). 
	The intensity profile of a Kerr lensed disk can be significantly different from that of an un-lensed disk. 
	Consequently, the half light radius of a Kerr lensed disk is different from un-lensed case.
	\label{fig:intensity_image}}
\end{figure}

\begin{figure}
	\epsscale{0.8}
	\plotone{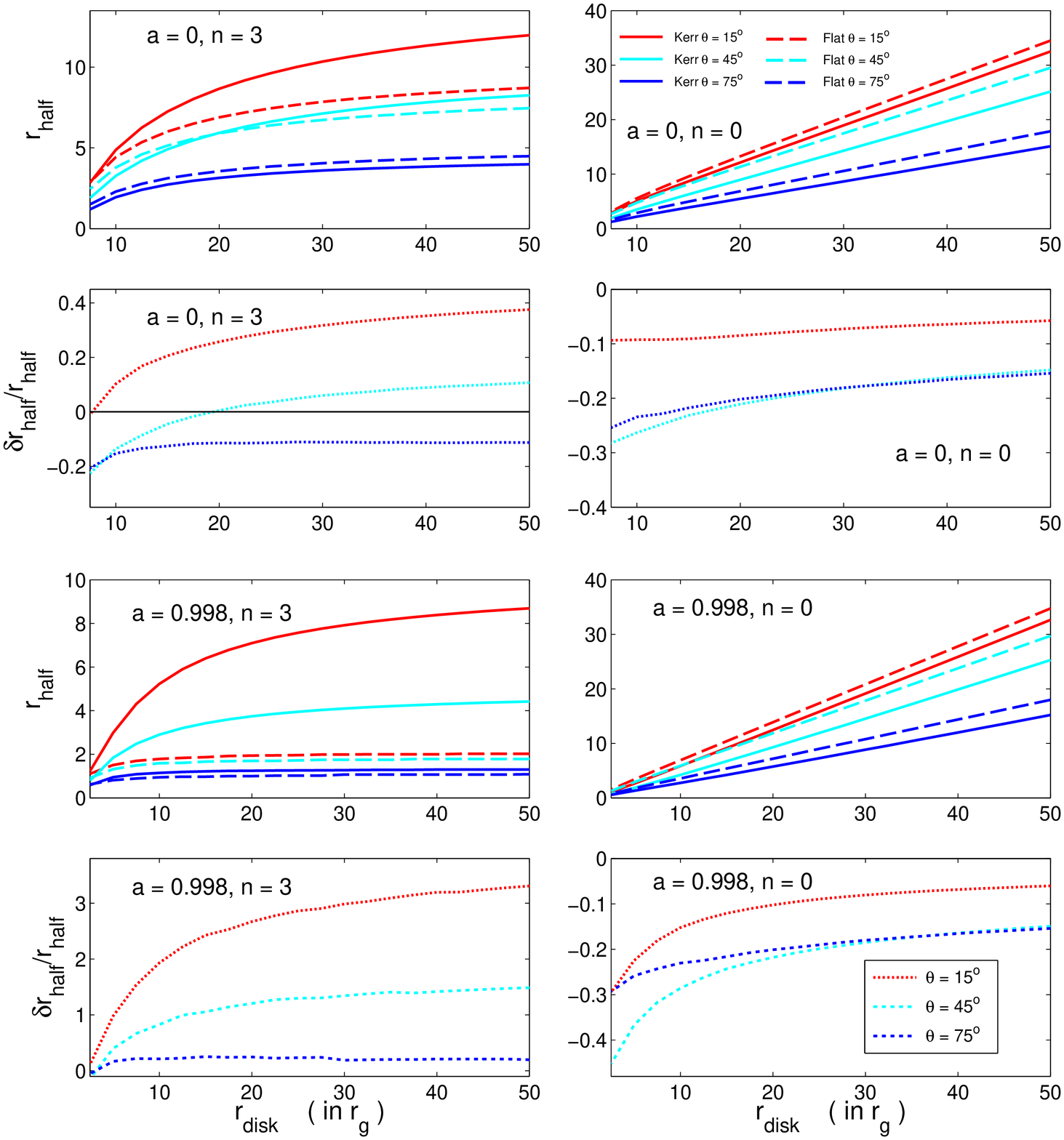}
	\caption{Strong lensing correction to half light radii of X-ray disks. 
	We plot $r_{\rm half}$ as a function of the emission size $r_{\rm disk}.$ 
	We show results for four source models  
	and for three inclination angles ($\theta = 15^{\circ},$ $45^\circ$ and $75^\circ$, as red, cyan, and blue curves).
	The data was given Table~\ref{tab:hl_radius}.
	The solid and dashed curves show lensed and un-lensed $r_{\rm half},$ respectively.
	The relative correction $(r_{\rm half}^{\rm Kerr}-r_{\rm half}^{\rm flat})/r_{\rm half}^{\rm flat}$  is shown in the second and fourth rows. 
	Relativistic effects change $r_{\rm half}$ significantly.
	For a flat radial profile ($n=0$, the second column) the lensed disk has smaller $r_{\rm half}$ than un-lensed disk (the relativistic beaming beats the area amplification caused by gravitational light bending, and gravitational redshift effect). 
	The reduction in $r_{\rm half}$ is about 5--40\% depending on the spin and inclination angle. 
	For a steep radial profile $n=3$ with $a=0.998\, M_{\rm BH},$ the dominating gravitational redshift effect makes the intensity profile much less concentrated toward the center, in particular for nearly face on case. 
	Consequently, $r_{\rm half}^{\rm Kerr}$ can be $\sim$3 times larger than $r_{\rm half}^{\rm flat}$. 
	This number reduces to $\sim$20\% when the disk is observed near edge on ($\theta=75^\circ$).
		\label{fig:hl_radius}}
\end{figure}

Kerr lensing changes the size, the shape, and the intensity profile of quasar X-ray emission regions.
Consequently, the microlensing light curves (e.g., the amplitude of magnitude fluctuation, $m_{\rm max-min}$) are different for flat or Kerr-lensed disks. 
Current quasar microlensing observations might have slightly over or underestimated the X-ray emission sizes.
We now tentatively estimate this bias.
What microlensing observations really constrain is the half light radius $r_{\rm half}$, with weak dependence on the emission profiles (Mortonson et al.\ 2005). 
We compare the half light radii of  flat X-ray disks, $r_{\rm half}^{\rm flat},$ with those of Kerr-lensed disks, $r_{\rm half}^{\rm Kerr}.$
We show the results in Figures~\ref{fig:intensity_image} and \ref{fig:hl_radius}, and Table~\ref{tab:hl_radius}.
Half light radius $r_{\rm half}$ was originally defined for sources with spherical symmetry to be the radius of the circular disk centering at the peak surface brightness and containing half the total flux.
For non-spherically symmetric objects, half right radius can be similarly defined to be the ``effective" radius of the half light region (from peak brightness down to a fixed surface brightness). 
Figure~\ref{fig:intensity_image} shows the intensity profile, peak brightness location, and half light region of images of X-ray disks lensed by Kerr strong gravity.
We  consider the same four source models and three inclination angles as before.  
The $r_{\rm half}$ for $2.5\,r_g< r_{\rm disk}< 50\,r_g$  are given in Table~\ref{tab:hl_radius} and shown in Figure~\ref{fig:hl_radius} for different source models and inclination angles (solid and dashed curves are for Kerr and flat disks, respectively; red, cyan, and blue curves are for inclination angle $\theta= 15^\circ,$ $45^\circ$ and $75^\circ$).

Strong gravity and relativistic flows change $r_{\rm half}$ in a few respects. 
First, the gravitational light bending increases the image area and tends to increase the half light radius.
This area distortion is more significant for high inclination angles (Chen et al.\ 2012b).
Second, the differential gravitational redshifts between different regions of the source emission change the intensity profiles of lensed disks.
This also changes $r_{\rm half}$.
For example,  the surface brightness distribution of an X-ray disk with steep radial profile ($n=3$) will be much less concentrated toward the center after the gravitational redshift effects have been taken into account.
Consequently, the gravitational redshift tend to increase the $r_{\rm half}$ for this case.
On the other hand, for an uniform intensity profile ($n=0$), differential gravitational redshifts focus the emission to regions less severely redshifted, this tends to reduces $r_{\rm half}$ (\ie $r_{\rm half}^{\rm Kerr}<\sqrt{{\rm A}/2\pi}=r_{\rm half}^{\rm flat}$ where $A$ is the disk area).   
Thirdly, the Doppler shifts and relativistic beaming tend to focus the intensity to a small region where the emission is strongly blueshifted. 
If the un-lensed emission is already concentrated in this region, the lensed profile will then be even more concentrated,  resulting in a smaller $r_{\rm half}$ value.
If not, this lensing produced (local) peak brightness location might compete with other regions of intrinsically strong emission, and can possibly give a larger $r_{\rm half}.$ 
Doppler shifts are more significant for high inclination angles (nearly edge on).  
In reality, $r_{\rm half}$ is influenced by these effects simultaneously.
For example, for $n=3,$ $a=0.998$, $\theta=15^\circ,$ the gravitational redshift effect dominates.
Consequently, $r^{\rm Kerr}_{\rm half}$ can be $\sim$3 times larger than $r_{\rm half}^{\rm flat}$ for $r_{\rm disk}\gtrsim 20\,r_g$, see the (4,1) panel of Figure \ref{fig:hl_radius}.
For $n=0,$ $a=0$ or $0.998M_{\rm BH},$  $r_{\rm half}^{\rm Kerr}$ is smaller than $r_{\rm half}^{\rm flat}$  by 10--40\% depending on the spin and inclination angle (see the second column of Figure~\ref{fig:hl_radius}).
For $n=3$, $a=0$ case, $r_{\rm half}^{\rm Kerr}$ can be either smaller or larger than $r^{\rm flat}_{\rm half}$ (from $-20\%$ to $+40\%$) depending on the emission size and the inclination angle (see the (1,1) and (2,1) panel of Figure~\ref{fig:hl_radius}).

We now estimate the bias of X-ray emission size using the $m_{\rm max-min}$ metric.
Assuming some amplitude of magnitude of fluctuation, $m_{\rm max-min},$ is observed, we compute the emission size $r_{\rm disk}$ and half light radius $r_{\rm half}$ needed to produce the same $m_{\rm max-min}$ for both flat and Kerr-lensed disk models. 
The difference between these two cases measures the bias of current microlensing measurements.     
We consider the same four source  models and three inclination angles as before (see Figure~\ref{fig:max_min_metric}). 
For each case, we test a couple of $m_{\rm max-min}$ values (marked by thin black lines).
First we find the corresponding $r_{\rm disk}$ values using Table~\ref{tab:delta_m_metric}.
We  then compute  the half light radius $r_{\rm half}$ using Table~\ref{tab:hl_radius}.
The results are shown in Table~\ref{tab:bias}.
The relative corrections to $r_{\rm disk}$ and $r_{\rm half}$ are given in the $4^{\rm th}$ and the $7^{\rm th}$ column, respectively. 
For most of the cases we considered, we found an underestimate of the half light radius (up to $\sim$50\%).
For example, we consider the $n=0,$ $a=0$ model with $m_{\rm max-min}=2.0$ (see the top right panel of Figure~\ref{fig:max_min_metric}). 
Assuming flat X-ray disks results in an underestimate of $r_{\rm disk}$ by $20\%$, $40\%$ and $49\%,$ and an underestimate of $r_{\rm half}$ by $18\%$, $33\%$ and $45\%$ for inclination angle  $\theta= 15^\circ,$ $45^\circ,$ and $75^\circ,$ respectively.
A larger emission size is needed for Kerr-lensed disk because for a fixed $r_{\rm disk},$ a Kerr disk has a smaller $r_{\rm half}$ (see the (2,2) panel of Figure~\ref{fig:hl_radius}) and this gives larger $m_{\rm max-min}.$
In order to produce the same $m_{\rm max-min}$ as a flat disk, we need to increase the source size for Kerr-lensed disks.
We found an overestimate of $r_{\rm half}$ only  for the $n=0,$ $a=0.998M_{\rm BH}$ and $\theta=15^\circ$ case, \ie a disk with flat radial profile, large spin, and observed nearly face on. 
For this model, an overestimate of $r_{\rm half}$ by $\sim$20\% is possible when the disk size is small, $r_{\rm disk}\lesssim 6\,r_g.$
For the case $a=0.998M_{\rm BH},$ $n=3,$ and $\theta=15^\circ,$ if we observe $m_{\rm max-min}=2.8,$ then an emission size $r_{\rm disk}^{\rm flat}=14.8\,r_g$ is needed assuming a flat disk,  $\sim$3 times larger than $r_{\rm disk}^{\rm Kerr}=4.0\,r_g$ assuming a Kerr-lensed disk (see the (2,1) panel of Figure~\ref{fig:max_min_metric}).
This is because the severe gravitational redshift suffered by the central regions of  Kerr disk makes the lensed intensity profile much less concentrated than the intrinsically steep radial profile, and therefore, a smaller $r_{\rm disk}^{\rm Kerr}$ (but much less concentrated) can produce similar amplitude of magnitude fluctuation $m_{\rm max-min}$ as a larger but more concentrated  flat disk  does.
In terms of half light radius (what microlensing really constrains),  we still found an underestimate ($\sim$19\%). 
This underestimate in $r_{\rm half}$ does not contradict the overestimate in $r_{\rm disk}.$ 
For a fixed $r_{\rm disk}$, an un-lensed disk has much smaller half light radius than a Kerr-lensed disk (see the dashed and solid red curves in the (3,1) panel of Figure~\ref{fig:hl_radius}).
Consequently, the half light radius of a flat disk with $r_{\rm disk}=14.8\,r_g$ is smaller than that of a lensed disk with $r_{\rm disk}=4.0\,r_g.$ 


Before ending this section, we  point out that  Kerr lensing can change the caustic crossing time significantly.
This can be easily seen for the simple Chang-Refsdal lens model (Figure~\ref{fig:LC_Chang}).
For example, for the case with inclination angle $\theta=75^\circ$, $a=0.998\,M_{\rm BH}$ and steep profile (n=3) (bottom left panel of Figure~\ref{fig:LC_Chang}), the two caustic crossing events of Kerr+Micro lensing light curves are later than those of standard microlensing light curves by $\sim$3.8 and 7.0 months respectively (assuming that the relative crossing velocity is $v_s=300\,\rm km\,s^{-1}$ in the source plane, see Mosquera \& Kochanek 2011), presumably caused by the focusing of emission to the approaching side of the disk (the left side, Figure~\ref{fig:mag_pattern}). 
A simultaneous monitoring of gravitational lensed quasars in both X-ray and optical bands with densely sampled X-ray light curves might reveal this feature because the effect of strong gravity is much less in the optical band (see Chen et al.\ 2012b).

\section{Conclusion}

\begin{figure}
	\epsscale{0.75}
	\plotone{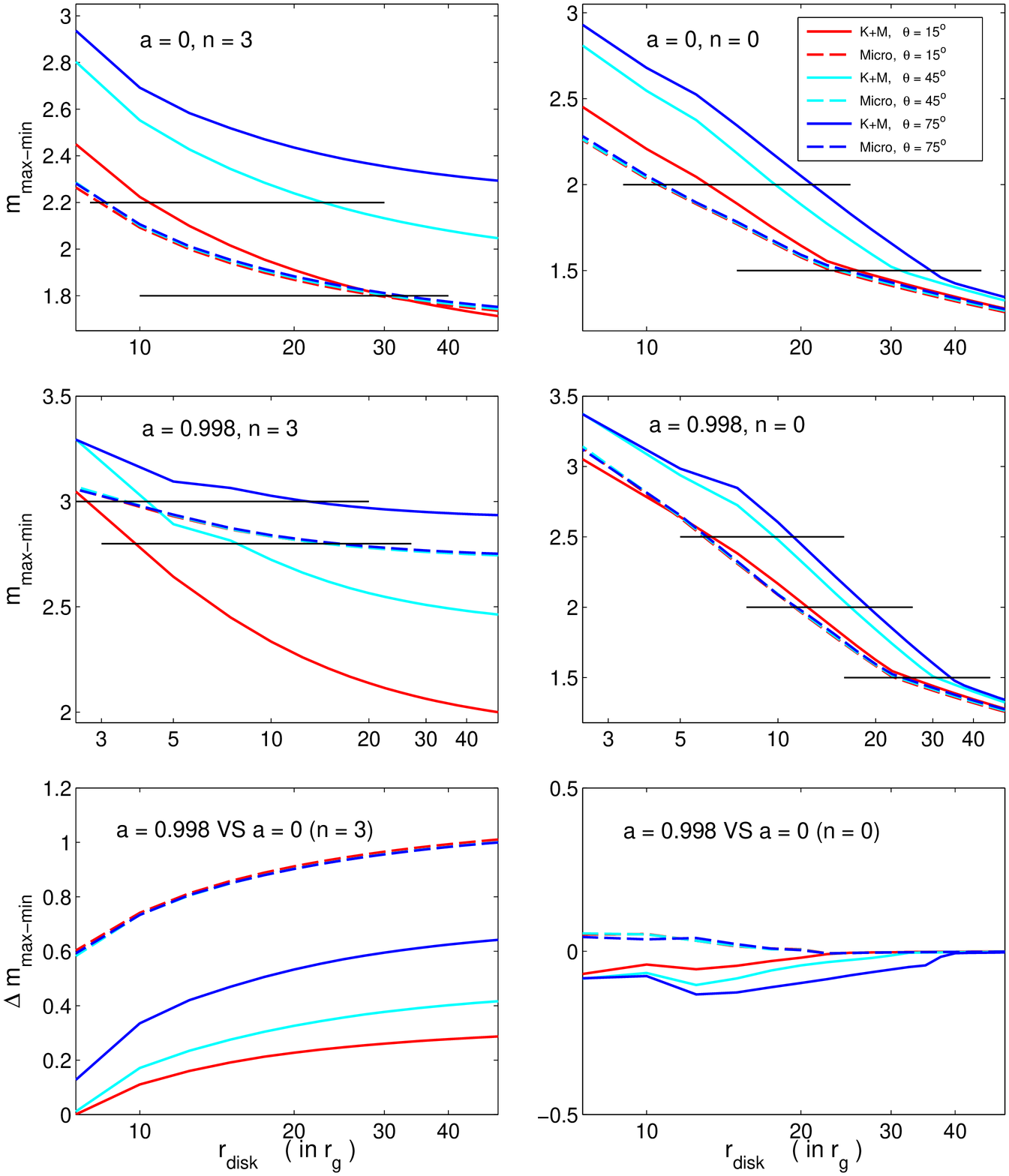}
	\caption{ $m_{\rm max-min}$ as a function of emission region size $r_{\rm disk}.$
	The microlensing magnification pattern and the source trajectory are shown in the right panel of Figure~\ref{fig:mag_pattern}. 
	We show results for 4 source models (spin $a =0$ or $0.998\,M_{\rm BH}$ and radial profile $n=3$ or $0$) and for three inclination angles $\theta=15^\circ,$ $45^\circ$ and $75^\circ$ (red, cyan, and blue curves) assuming flat or Kerr lensed disks (dashed and solid curves).
	The inclination angle dependence of $m_{\rm max-min}$ is much more significant for Kerr-lensed disks. 
	The horizontal lines in the first two rows mark the values of $m_{\rm max-min}$ which we used to estimate the bias of size constraints of current microlensing observations ignoring Kerr strong gravity.
	The spin-dependence of $m_{\rm max-min},$ \ie $m_{\rm max-min}^{(a=0.998)}-m_{\rm max-min}^{(a=0)}$ is shown in the third row.	
	For $n=3$ case, the spin dependence of $m_{\rm max-min}$ is more significant for a flat X-ray disk because the extra emission between $1.24\,r_g<r<6\,r_g$ is washed out by gravitational redshift for Kerr lensed X-ray disks.
	For $n=0$ case, the spin dependence of $m_{\rm max-min}$ for both flat and Kerr disks is insignificant since the extra emission between $1.24\,r_g<r<6\,r_g$, gravitationally redshifted or not, is not important for a flat 
	radial profile.
			\label{fig:max_min_metric}}
\end{figure}

We have combined Kerr lensing and standard microlensing to produce what we call Kerr+Micro lensing light curves.
We have studied the effect of Kerr lensing on microlensing observations using a simple X-ray source geometry and two simple models for the foreground microlensing.
The strong lensing of the X-ray emission by the central supermassive black hole  changes the size, shape, and profile of the original X-ray emission.
Kerr strong gravity changes the observed quasar X-ray microlensing light curves in a nontrivial way.
Kerr+Micro lensing light curves have a larger amplitude of magnitude fluctuation for larger inclination angle. 
In particular, Kerr lensing can reduce or increase the amplitude of the magnitude fluctuation of microlensing light curves by $\sim$0.65--0.75 mag depending on the inclination angle of the observer, the spin of the black hole, and the intensity profile of the corona (see Figure~\ref{fig:max_min_metric}).
Consequently, current quasar microlensing observations might have over or underestimated the X-ray emission sizes. 
We estimate this bias using a simple metric base on the amplitude of magnitude fluctuation (refer to \textsection\ref{sec:hl_radius}). 
For most of the cases we considered,  we found an underestimate of the half light radius (up to $\sim$50\%).
An overestimate of half light radius (up to $\sim$20\%) was only found for the $a=0.998$, $n=0,$ $\theta=15^\circ$ case (Table~\ref{tab:bias}).
We conclude that current microlensing size measurements generally underestimate the true size of the X-ray emission regions, and that more accurate constraints  should be obtainable from microlenisng observations by including Kerr lensing.
Furthermore, it should be possible to measure the inclination angle and the black hole spin by analyzing Kerr+Micro lensing light curves.
This new lensing model might help in breaking the parameter degeneracy of other methods measuring AGN black hole spin and/or inclination angle, e.g.,  by broad \feka\ line shape (Reynolds \& Fabian 2008; de La Calle P$\rm\acute{e}$rez et al.\ 2010), continuum fitting (Davis et al.\ 2006; Shafee et al.\ 2006; Czerny et al.\ 2011), polarization of the continuum emission (Dov$\rm\check{c}$iak et al.\ 2008; Li et al.\ 2009; Schnittman \& Krolik 2009), and high frequency quasi-periodic oscillations (T$\rm\ddot{o}$r$\rm\ddot{o}$k et al.\ 2005; Middleton et al.\ 2011; Das \& Czerny 2011).

Applications to other X-ray models such as an X-ray ball or an X-ray disk above the black hole  (Chen et al.\ 2012b) are straightforward and we expect the results to be similar provided that the X-ray source is of a similar size to those assumed here, and that they are close to the central black hole.
We have chosen a cutoff of the X-ray emission at the innermost stable circular orbit.
In principle, it is possible to extend the X-ray emission region to within $r_{\rm ISCO}$ where the gas follows so-called plunging trajectories along the geodesics (Agol \& Krolik 2000).
We expect the effect of Kerr lensing will be more important for that case.
As a simple demonstration, we have only probed Kerr lensing effects by using two typical magnification patterns each with a single crossing path.
To fully characterize  the effects of Kerr lensing for microlensing applications, a detailed modeling of a known microlensed quasar with well-sampled light curves in the X-ray band, such as Q~2237+0305  (Chen et al.\ 2012a) and RXJ~1131$-$1231 (Chartas  et al.\ 2012) combining the double lensing technique introduced in this paper and Bayesian Monte-Carlo analysis (Kochanek 2004) will be pursued in future work.

We thank C. S. Kochanek for comments and suggestions.
BC and XD acknowledge support for this work provided by the National Aeronautics and Space Administration through Chandra Award Number GO0-11121B, GO1-12139B, GO2-13132A issued by the Chandra X-ray Observatory Center, which is operated by the Smithsonian Astrophysical Observatory for and on behalf of the National Aeronautics Space Administration under contract NAS8-03060.
BC and XD acknowledge support for program number HST-GO-11732.07-A  provided by NASA through a grant from the Space Telescope Science Institute, which is operated by the Association of Universities for Research in Astronomy, Incorporated.
BC and EB acknowledge NSF AST-0707704, and US DOE Grant DE-FG02-07ER41517 and support for program number HST-GO-12298.05-A provided by NASA through a grant from the Space Telescope Science Institute, which is operated by the Association of Universities for Research in Astronomy, Incorporated, under NASA contract NAS5-26555.

\begin{deluxetable}{cllllllll}
\tabletypesize{\scriptsize}
\tablecolumns{9}
\tablewidth{0pt}
\tablecaption{Magnitude fluctuation ($m_{\rm max-min}$) of the light curves of Kerr+Micro lensing and  microlensing. Source size $r_{\rm disk}=20\,r_g.$  \label{tab:dMag}}
\tablehead{
\colhead{Lens\tablenotemark{a}}&
\colhead{Source}&
\multicolumn{3}{c}{Microlensing} & \colhead{} & \multicolumn{3}{c}{Kerr+Micro Lensing\tablenotemark{b}}\\
\cline{3-5}\cline{7-9}\\
\colhead{} &
\colhead{}&
\colhead{ $\theta=15^\circ$}&
\colhead{ $\theta=45^\circ$}&
\colhead{ $\theta=75^\circ$}&
\colhead{}&
\colhead{ $\theta=15^\circ$}&
\colhead{ $\theta=45^\circ$}&
\colhead{ $\theta=75^\circ$}
}
\startdata
I & ${\rm  a=0.998,n=3}     $ &  2.42 &  2.41 &   2.42  &  & 1.96 ($-$0.46)      &  2.28 ($-$0.13)      &  2.76 ($+$0.33)    \\
I & ${\rm  a=0.998,n=0}     $ & 1.70  & 1.70 &   1.71   &  & 1.75 ($+$0.05)      & 1.85 ($+$0.14)      & 1.86 ($+$0.16)       \\
I & ${\rm  a=0, \>\>\>\>\>\>\>\> n=3}            $ & 1.88  & 1.88  &  1.88  &   & 1.86 ($-$0.02)       &  2.00 ($+$0.12)     &  2.19 ($+$0.31)     \\
I & ${\rm  a=0, \>\>\>\>\>\>\>\>n=0}            $ & 1.71  & 1.71  &  1.71  &   & 1.76 ($+$0.05)      & 1.86 ($+$0.15)     &  1.90 ($+$0.18)     \\
 \tableline
 II & ${\rm  a=0.998,n=3}    $ &  2.78  &  2.78 &  2.79 &    & 2.14 ($-$0.64)      &  2.56 ($-$0.22)      &   2.98 ($+$0.19)    \\
II & ${\rm  a=0.998,n=0}    $  &  1.58  & 1.59  & 1.60  &     &1.63 ($+$0.04)      & 1.84 ($+$0.25)      &   1.95 ($+$0.36)       \\
II & ${\rm  a=0, \>\>\>\>\>\>\>\>n=3}           $  & 1.87  & 1.88   & 1.88  &     & 1.91 ($+$0.04)      &  2.24 ($+$0.37)    &  2.44 ($+$0.55)     \\
II & ${\rm  a=0, \>\>\>\>\>\>\>\>n=0}           $  & 1.58  & 1.59   & 1.59  &     & 1.65 ($+$0.07)     & 1.89 ($+$0.30)      &  2.05 ($+$0.46)     \\
\enddata
\tablenotetext{a}{I: Chang-Refsdal lens model; II: Random star field.}
\tablenotetext{b}{The number in the parenthesis is the differential magnitude fluctuation between Kerr+Micro lensing and microlensing for same inclination angle $\theta$, \ie 
$m_{\rm max-min}^{\rm K+M}(\theta)-m_{\rm max-min}^{\rm Micro}(\theta)$.}
\end{deluxetable}


\begin{deluxetable}{lllllllllllll}
\tabletypesize{\scriptsize}
\tablecolumns{13}
\tablewidth{0pt}
\tablecaption{Lensed/un-lensed half light radius $r_{\rm half}$ of X-ray disk as a function of the emission size $r_{\rm disk}$. 
\label{tab:hl_radius}}
\tablehead{
\colhead{Model\tablenotemark{a}}&
\multicolumn{11}{c}{$r_{\rm disk}$} \\
\cline{2-12} \\
\colhead{}&
\colhead{ $50\,r_g$}&
\colhead{ $45\,r_g$}&
\colhead{ $40\,r_g$}&
\colhead{ $35\,r_g$} &
\colhead{ $30\,r_g$}&
\colhead{ $25\,r_g$}&
\colhead{ $20\,r_g$} &
\colhead{$15\,r_g$} &
\colhead{ $10\,r_g$}&
\colhead{ $5\,r_g$\tablenotemark{b}}&
\colhead{ $2.5\,r_g$\tablenotemark{b}}
}
\startdata
Kerr Disk   &    &     &   &     &    &      &        &     &  &   &      \\
\tableline
$a=0, \>\>\>\>\>\>\>\>\>n=3,\>\theta=15^\circ$ & 11.98&11.68&11.33&10.89&10.34& 9.63& 8.66& 7.25& 4.88	& \nodata & \nodata \\
$a=0, \>\>\>\>\>\>\>\>\>n=3,\>\theta=45^\circ$ & 8.26& 8.06& 7.82& 7.51& 7.13& 6.62& 5.93& 4.91& 3.27	& \nodata & \nodata \\
$a=0, \>\>\>\>\>\>\>\>\>n=3,\>\theta=75^\circ$ & 3.98& 3.92& 3.83& 3.73& 3.59& 3.41& 3.14& 2.72& 1.94	& \nodata & \nodata \\
$a=0.998, \>n=3,\>\theta=15^\circ$ & 8.69& 8.55& 8.39& 8.18& 7.92& 7.57& 7.10& 6.40& 5.23			& 2.98       & 1.23  \\
$a=0.998, \>n=3,\>\theta=45^\circ$ & 4.42& 4.36& 4.29& 4.21& 4.09& 3.95& 3.74& 3.42& 2.90			& 1.84      &  0.79 \\
$a=0.998, \>n=3,\>\theta=75^\circ$ & 1.30& 1.30& 1.30& 1.29& 1.27& 1.26& 1.24& 1.22& 1.15			& 0.95      &  0.59 \\
$a=0, \>\>\>\>\>\>\>\>\>n=0,\>\theta=15^\circ$ & 32.51&29.11&25.72&22.33&18.94&15.54&12.14& 8.69& 5.05& \nodata & \nodata \\
$a=0,\>\>\>\>\>\>\>\>\>n=0,\>\theta=45^\circ$ & 25.14&22.41&19.70&16.99&14.30&11.62& 8.95& 6.28& 3.51	  & \nodata & \nodata \\
$a=0, \>\>\>\>\>\>\>\>\>n=0,\>\theta=75^\circ$ & 15.10&13.48&11.87&10.26& 8.66& 7.07& 5.48& 3.88& 2.20    & \nodata & \nodata \\
$a=0.998, \>n=0,\>\theta=15^\circ$ & 32.64&29.25&25.88&22.51&19.15&15.79&12.45& 9.13& 5.85& 2.61        & 1.08		\\
$a=0.998, \>n=0,\>\theta=45^\circ$ & 25.28&22.57&19.87&17.19&14.53&11.89& 9.28& 6.72& 4.21& 1.82	   & 0.70		\\
$a=0.998, \>n=0,\>\theta=75^\circ$ &15.21&13.60&12.00&10.41& 8.83& 7.27& 5.73& 4.22& 2.74& 1.30	            & 0.55		\\
\tableline
Flat Disk   &    &     &   &     &    &      &        &     &  &   &      \\
\tableline
$a=0, \>\>\>\>\>\>\>\>\>n=3,\>\theta=15^\circ$ & 8.71& 8.56& 8.38& 8.15& 7.85& 7.45& 6.89& 6.01& 4.42& \nodata & \nodata\\
$a=0, \>\>\>\>\>\>\>\>\>n=3,\>\theta=45^\circ$ & 7.46& 7.34& 7.18& 6.99& 6.72& 6.40& 5.90& 5.14& 3.78& \nodata & \nodata\\
$a=0, \>\>\>\>\>\>\>\>\>n=3,\>\theta=75^\circ$ & 4.49& 4.41& 4.32& 4.21& 4.05& 3.85& 3.55& 3.11& 2.29& \nodata &\nodata \\
$a=0.998, \>n=3,\>\theta=15^\circ$ & 2.02& 2.02& 2.00& 2.00& 1.99& 1.96& 1.93& 1.87& 1.78& 1.51   &    1.11	\\
$a=0.998, \>n=3,\>\theta=45^\circ$ & 1.78& 1.78& 1.78& 1.75& 1.75& 1.72& 1.70& 1.67& 1.59& 1.32   &	   0.92	\\
$a=0.998, \>n=3,\>\theta=75^\circ$ & 1.08& 1.07& 1.07& 1.07& 1.07& 1.02& 1.00& 0.97& 0.94& 0.81   &   0.62 	\\
$a=0, \>\>\>\>\>\>\>\>\>n=0,\>\theta=15^\circ$ & 34.50&30.99&27.49&23.96&20.43&16.86&13.26& 9.56& 5.57& \nodata & \nodata\\
$a=0, \>\>\>\>\>\>\>\>\>n=0,\>\theta=45^\circ$ & 29.51&26.52&23.51&20.50&17.48&14.43&11.34& 8.17& 4.76& \nodata & \nodata \\
$a=0, \>\>\>\>\>\>\>\>\>n=0,\>\theta=75^\circ$ & 17.86&16.05&14.23&12.41&10.57& 8.74& 6.86& 4.96& 2.88& \nodata    &\nodata\\
$a=0.998, \>n=0,\>\theta=15^\circ$ & 34.74&31.26&27.79&24.31&20.83&17.35&13.87&10.39& 6.90& 3.37  & 1.53 \\
$a=0.998, \>n=0,\>\theta=45^\circ$ & 29.72&26.75&23.77&20.80&17.82&14.85&11.87& 8.88& 5.90& 2.87   & 1.29 \\
$a=0.998, \>n=0,\>\theta=75^\circ$ & 17.98&16.18&14.38&12.58&10.77& 8.99& 7.17& 5.38& 3.56& 1.75      & 0.78 \\
\enddata
\tablenotetext{a}{The images and half light regions for a Kerr lensed disk with $r_{\rm disk}=50\,r_g$ are shown in Figure~\ref{fig:intensity_image}.
			The half light radius as a function of disk size $r_{\rm disk}$ is shown in Figure~\ref{fig:hl_radius}.}
\tablenotetext{b}{For $a=0$ case, since $r_{\rm ISCO}= 6\,r_g$, the X-ray disk model is not defined for these two cases.  }
\end{deluxetable}


\begin{deluxetable}{lllllllllllll}
\tabletypesize{\scriptsize}
\tablecolumns{13}
\tablewidth{0pt}
\tablecaption{Amplitude of magnitude fluctuation, $m_{\rm max-min},$ of  light curves of X-ray disks as a function of the emission size $r_{\rm disk}$. 
\label{tab:delta_m_metric}}
\tablehead{
\colhead{Model}&
\multicolumn{11}{c}{$r_{\rm disk}$} \\
\cline{2-12} \\
\colhead{}&
\colhead{ $50\,r_g$}&
\colhead{ $45\,r_g$}&
\colhead{ $40\,r_g$}&
\colhead{ $35\,r_g$} &
\colhead{ $30\,r_g$}&
\colhead{ $25\,r_g$}&
\colhead{ $20\,r_g$} &
\colhead{$15\,r_g$} &
\colhead{ $10\,r_g$}&
\colhead{ $5\,r_g$\tablenotemark{b}}&
\colhead{ $2.5\,r_g$\tablenotemark{b}}
}
\startdata
Kerr Disk   &    &     &   &     &    &      &        &     &  &   &  &    \\
\tableline
$a=0, \>\>\>\>\>\>\>\>\>n=3,\>\theta=15^\circ$ & 1.71& 1.73& 1.75& 1.77& 1.80& 1.85& 1.91& 2.02& 2.22 & \nodata &\nodata  \\
$a=0, \>\>\>\>\>\>\>\>\>n=3,\>\theta=45^\circ$ &  2.05& 2.06& 2.08& 2.10& 2.13& 2.18& 2.24& 2.34& 2.55& \nodata &\nodata  \\
$a=0, \>\>\>\>\>\>\>\>\>n=3,\>\theta=75^\circ$ &  2.29& 2.30& 2.32& 2.33& 2.35& 2.39& 2.44& 2.52& 2.69& \nodata &\nodata  \\
$a=0.998, \>n=3,\>\theta=15^\circ$                    &  2.00& 2.01& 2.02& 2.04& 2.06& 2.09& 2.14& 2.21& 2.33& 2.64       & 3.05 \\
$a=0.998, \>n=3,\>\theta=45^\circ$                    &  2.46& 2.47& 2.48& 2.49& 2.51& 2.53& 2.57& 2.62& 2.72& 2.89       & 3.29 \\
$a=0.998, \>n=3,\>\theta=75^\circ$                    &  2.93& 2.94& 2.94& 2.94& 2.95& 2.96& 2.97& 2.99& 3.03& 3.09       & 3.29 \\
$a=0, \>\>\>\>\>\>\>\>\>n=0,\>\theta=15^\circ$ & 1.28& 1.31& 1.35& 1.39& 1.44& 1.51& 1.65& 1.89& 2.21& \nodata   & \nodata \\
$a=0, \>\>\>\>\>\>\>\>\>n=0,\>\theta=45^\circ$ & 1.33& 1.36& 1.40& 1.45& 1.52& 1.68& 1.89& 2.18& 2.55& \nodata   & \nodata\\
$a=0, \>\>\>\>\>\>\>\>\>n=0,\>\theta=75^\circ$ & 1.35& 1.38& 1.43& 1.52& 1.66& 1.83& 2.05& 2.35& 2.68& \nodata   & \nodata\\
$a=0.998, \>n=0,\>\theta=15^\circ$                    & 1.28& 1.31& 1.35& 1.39& 1.44& 1.51& 1.63& 1.85& 2.17& 2.64         & 3.05  \\
$a=0.998, \>n=0,\>\theta=45^\circ$                    & 1.32& 1.36& 1.40& 1.45& 1.51& 1.65& 1.84& 2.10& 2.48& 2.94         & 3.37 \\
$a=0.998, \>n=0,\>\theta=75^\circ$                    & 1.34& 1.38& 1.42& 1.48& 1.60& 1.76& 1.95& 2.22& 2.60& 2.98         & 3.37 \\
\tableline
Flat Disk   &    &     &   &     &    &      &        &     &  &   &    &  \\
\tableline  
$a=0, \>\>\>\>\>\>\>\>\>n=3,\>\theta=15^\circ$ & 1.74& 1.75& 1.76& 1.77& 1.80& 1.83& 1.87& 1.94& 2.09& \nodata  & \nodata\\
$a=0, \>\>\>\>\>\>\>\>\>n=3,\>\theta=45^\circ$ & 1.74& 1.75& 1.77& 1.78& 1.80& 1.83& 1.88& 1.95& 2.10& \nodata  & \nodata\\
$a=0, \>\>\>\>\>\>\>\>\>n=3,\>\theta=75^\circ$ & 1.75& 1.76& 1.77& 1.79& 1.81& 1.84& 1.88& 1.95& 2.11& \nodata  & \nodata \\
$a=0.998, \>n=3,\>\theta=15^\circ$                    & 2.75& 2.75& 2.75& 2.76& 2.76& 2.77& 2.78& 2.80& 2.83& 2.93       & 3.06 \\
$a=0.998, \>n=3,\>\theta=45^\circ$                    & 2.74& 2.75& 2.75& 2.75& 2.76& 2.77& 2.78& 2.80& 2.83& 2.93       & 3.07 \\
$a=0.998, \>n=3,\>\theta=75^\circ$                    & 2.75& 2.75& 2.76& 2.76& 2.77& 2.78& 2.79& 2.80& 2.84& 2.94       & 3.07\\
$a=0, \>\>\>\>\>\>\>\>\>n=0,\>\theta=15^\circ$ & 1.26& 1.29& 1.32& 1.36& 1.41& 1.47& 1.58& 1.77& 2.03& \nodata  & \nodata \\
$a=0, \>\>\>\>\>\>\>\>\>n=0,\>\theta=45^\circ$ & 1.27& 1.30& 1.33& 1.37& 1.42& 1.48& 1.59& 1.78& 2.04& \nodata  & \nodata \\
$a=0, \>\>\>\>\>\>\>\>\>n=0,\>\theta=75^\circ$ & 1.28& 1.31& 1.34& 1.38& 1.43& 1.49& 1.59& 1.78& 2.05& \nodata  & \nodata\\
$a=0.998, \>n=0,\>\theta=15^\circ$                    & 1.26& 1.29& 1.32& 1.36& 1.41& 1.47& 1.58& 1.79& 2.09& 2.64       & 3.13 \\
$a=0.998, \>n=0,\>\theta=45^\circ$                    & 1.27& 1.29& 1.33& 1.37& 1.42& 1.48& 1.59& 1.79& 2.09& 2.64      &  3.14\\
$a=0.998, \>n=0,\>\theta=75^\circ$                    & 1.27& 1.30& 1.34& 1.38& 1.43& 1.49& 1.60& 1.81& 2.09& 2.65      &  3.14 \\
\enddata
\tablenotetext{b}{For $a=0$ case, since $r_{\rm ISCO}= 6\,r_g$, the X-ray disk model is not defined for these two cases.  }
\end{deluxetable}


\begin{deluxetable}{lllllll}
\tabletypesize{\scriptsize}
\tablecolumns{7}
\tablewidth{0pt}
\tablecaption{Bias estimate of microlensing size constraints ignoring Kerr strong gravity. 
\label{tab:bias}}
\tablehead{
\colhead{Model\tablenotemark{a}} &
\colhead{$r_{\rm disk}^{\rm flat}  $\tablenotemark{b}} & 
\colhead{$r_{\rm disk}^{\rm Kerr}\>$\tablenotemark{b}} &
\colhead{$\frac{\delta r_{\rm disk}}{r_{\rm disk}^{\rm Kerr}}$\tablenotemark{b}} &
\colhead{$r_{\rm half}^{\rm flat}  $} & 
\colhead{$r_{\rm half}^{\rm Kerr}$} &
\colhead{$\frac{\delta r_{\rm half}}{r_{\rm half}^{\rm Kerr}}$} 
}
\startdata
$a=0,n=3$   &    &     &   &     &    &            \\
\tableline
$\theta=15^\circ,\>m_{\rm max-min}=2.2$ & 8.43 & 10.48    & $-19.5\%$ & 3.45 & 5.14      & $-32.9\%$ \\
$\theta=45^\circ,\>m_{\rm max-min}=2.2$ & 8.66 & 22.81    & $-62.1\%$ & 3.07 & 6.34      & $-51.6\%$ \\
$\theta=75^\circ,\>m_{\rm max-min}=2.2$ & 9.66 & \nodata\tablenotemark{c} & \nodata      & 2.18 & \nodata & \nodata \\
\tableline
$\theta=15^\circ,\>m_{\rm max-min}=1.8$ & 29.2 & 30.45    & $-3.98\%$ & 7.79   & 10.39    & $-25.0\%$ \\
$\theta=45^\circ,\>m_{\rm max-min}=1.8$ & 30.8 & \nodata & \nodata      & 6.77   & \nodata & \nodata  \\
$\theta=75^\circ,\>m_{\rm max-min}=1.8$ & 32.6 & \nodata & \nodata      & 4.13   & \nodata & \nodata  \\
\tableline
$a=0,n=0$   &    &     &   &     &    &            \\
\tableline
$\theta=15^\circ,\>m_{\rm max-min}=2.0$ & 10.6 & 13.2 & $-20.0\%$ & 6.06 & 7.41 & $-18.3\%$ \\
$\theta=45^\circ,\>m_{\rm max-min}=2.0$ & 10.7 & 17.9 & $-40.1\%$ & 5.26 & 7.83 & $-32.9\%$ \\
$\theta=75^\circ,\>m_{\rm max-min}=2.0$ & 10.8 & 21.1 & $-48.6\%$ & 3.22 & 5.83 & $-44.7\%$ \\
\tableline
$\theta=15^\circ,\>m_{\rm max-min}=1.5$ & 23.3 & 25.8 & $-9.87\%$ & 15.62 & 16.09 & $-2.92\%$ \\
$\theta=45^\circ,\>m_{\rm max-min}=1.5$ & 23.9 & 31.4 & $-23.9\%$ & 13.77 & 15.06 & $-8.57\%$ \\
$\theta=75^\circ,\>m_{\rm max-min}=1.5$ & 24.6 & 35.9 & $-31.4\%$ & 8.59   & 10.54 & $-18.5\%$ \\
\tableline
$a=0.998,n=3$   &    &     &   &     &    &          \\
\tableline
$\theta=15^\circ,\>m_{\rm max-min}=3.0$ & 3.65 & 2.79   & $+30.9\%$ & 1.29 & 1.43  & $-10.0\%$\\
$\theta=45^\circ,\>m_{\rm max-min}=3.0$ & 3.79 & 4.33   & $-12.5\%$ &  1.12 & 1.56  & $-28.0\%$\\
$\theta=75^\circ,\>m_{\rm max-min}=3.0$ & 3.70 & 13.1   & $-71.7\%$ &  0.71 & 1.19  & $-40.4\%$\\
\tableline
$\theta=15^\circ,\>m_{\rm max-min}=2.8$ & 14.8 & 4.03       & $+267\%$  & 1.86   & 2.30      & $-18.9\%$\\
$\theta=45^\circ,\>m_{\rm max-min}=2.8$ & 14.7 & 7.89       & $+86.3\%$ & 1.66   & 2.55      & $-34.9\%$\\
$\theta=75^\circ,\>m_{\rm max-min}=2.8$ & 16.2 & \nodata & \nodata        & 0.98  & \nodata & \nodata  \\
\tableline
$a=0.998, n=0$   &    &     &   &     &    &            \\
\tableline
$\theta=15^\circ,\>m_{\rm max-min}=2.5$ & 6.04 & 6.37     & $-5.2\%$     & 4.11 & 3.49 & $+17.7\%$\\
$\theta=45^\circ,\>m_{\rm max-min}=2.5$ & 6.11 & 9.79     & $-37.7\%$   & 3.54 & 4.11 & $-13.8\%$\\
$\theta=75^\circ,\>m_{\rm max-min}=2.5$ & 6.17 & 11.23   & $-45.0\%$   & 2.18 & 3.10 & $-29.8\%$\\
\tableline
$\theta=15^\circ,\>m_{\rm max-min}=2.0$ & 11.31 & 12.36   & $-8.53\%$ & 7.81 & 7.39 & $+5.7\%$\\
$\theta=45^\circ,\>m_{\rm max-min}=2.0$ & 11.40 & 16.79   & $-32.1\%$ & 6.74 & 7.63 & $-11.7\%$\\
$\theta=75^\circ,\>m_{\rm max-min}=2.0$ & 11.47 & 19.07   & $-39.8\%$ & 4.10 & 5.45 & $-24.7\%$\\
\tableline
$\theta=15^\circ,\>m_{\rm max-min}=1.5$ & 22.86 & 25.50   & $-10.4\%$ & 15.86 & 16.13  & $-1.6\%$\\
$\theta=45^\circ,\>m_{\rm max-min}=1.5$ & 23.51 & 30.75   & $-23.6\%$ & 13.96 & 14.92  & $-6.5\%$\\
$\theta=75^\circ,\>m_{\rm max-min}=1.5$ & 24.24 & 34.08   & $-28.9\%$ &  8.71   & 10.12 & $-13.9\%$\\
\enddata
\tablenotetext{a}{We show results for 4 source models, spin $a =0$ or $0.998\,M_{\rm BH}$ and radial profile $n=3$ or $0,$
			and for three inclination angles $\theta = 15^{\circ},$ $45^\circ$ and $75^\circ$.
			For each model, the $m_{\rm max-min}$ values sampled are marked in the corresponding panels in Figure~\ref{fig:max_min_metric}.}
\tablenotetext{b}{These are less robust measurements from microlensing.}			
\tablenotetext{c}{There is no data because the $m_{\rm max-min}$ value sampled is not in the range of the $m_{\rm max-min}(r_{\rm disk})$ curve.
			  Refer to Figure~\ref{fig:max_min_metric}.  }
\end{deluxetable}

\end{document}